\documentclass[12pt]{article}
\usepackage{amsmath}
\usepackage{amssymb}
\usepackage{cancel}
\usepackage{graphicx}
\usepackage{color}
\usepackage{wrapfig}
\usepackage{arydshln}

\renewcommand{\thefootnote}{\fnsymbol{footnote}}

\begin{document}
\baselineskip=19.5pt

\begin{titlepage}
\begin{flushright}
{\small MISC-2014-08}
\end{flushright}
\begin{center}
\vspace*{17mm}

{\large\bf%
The spectrum and flavor composition of the astrophysical neutrinos in IceCube
}

\vspace*{10mm}
Atsushi~Watanabe
\footnote[1]{~watanabea@cc.kyoto-su.ac.jp}
\vspace*{10mm}

{\small {\it 
Maskawa Institute for Science and Culture, Kyoto Sangyo University, Kyoto 603-8555, Japan
}}\\

\vspace*{3mm}

{\small (December, 2014)}
\end{center}

\vspace*{7mm}

\begin{abstract}\noindent%
We fit the energy distribution of the IceCube starting events by a model which 
involves four parameters in the neutrino spectrum, namely three normalizations 
$n_e,n_\mu,n_\tau$ and a common power-law index $\gamma$, with a fixed background 
simulated by IceCube.
It is found that the best fit index is $\gamma = 2.7$ with $\chi^2_{\rm min} = 
32.3/24\,{\rm dof}$.
As for the two parameter model involving a democratic normalization and
an index, the best fit is at $\gamma = 2.8$ with $\chi^2_{\rm min} 
= 33.9/26\,{\rm dof}$.
The flavored model and the democratic model do not have much difference
in the quality of the (energy-spectrum) fit.
The standard $1:1:1$ composition is not disfavored by the current data.
\end{abstract}

\end{titlepage}

\newpage
\renewcommand{\thefootnote}{\fnsymbol{footnote}}
%%%%%%%%%%%%%%%%%%%%%%%%%%%%%%%%%%%%%%%%%%%%%%%%%%%%%%%%%%%%%%%%%%%%%%
\section{Introduction}
\label{intro}
%%%%%%%%%%%%%%%%%%%%%%%%%%%%%%%%%%%%%%%%%%%%%%%%%%%%%%%%%%%%%%%%%%%%%%
IceCube has recently made a great success in the observation of the high-energy 
neutrinos of extraterrestrial origin.
In their three years of data, 37 events have been found in $30$ TeV--$2$ PeV 
energy range~\cite{IC1,IC2,IC3}. 
They have concluded that the hypothesis of the atmospheric neutrino origin is 
rejected at 5.7$\sigma$, heralding a new era of high-energy astronomy.
The analysis with a lowered threshold down to $1$ TeV also shows a significant
contribution from the astrophysical component~\cite{IC4}.
Neutrino sky which will be seen by the existing and the future neutrino telescopes 
will provide indispensable information to understand the origin of cosmic rays, 
physics of Gamma Ray Bursts, GZK processes, etc.
Since the first announcement of the two PeV cascades, many authors have speculated
about the sources of the observed high-energy neutrinos~\cite{astro}.

While the high-energy neutrinos are unique astronomical messengers, they may 
also play an interesting role in particle physics.
Neutrino decay~\cite{decay}, pseudo-Dirac neutrinos~\cite{pDirac}, 
and Lorentz/CPT violation~\cite{CPT} 
have been discussed for long time as new physics testable by high-energy 
neutrinos.
More recently, the isolated nature of the events around $1$ PeV~\cite{IC3}
have triggered a variety of intriguing ideas such as 
decay of long-lived particles~\cite{PeVDM}, exotic mediators for neutrino 
absorbtions~\cite{abs}, new physics in the detection processes~\cite{detection}.
Obviously more data is needed to deal with such diverse hypotheses and speculations.

Flavor ratios for the three types of neutrinos are one of the key information for making 
further progress in these subjects~\cite{FR1}.
One of the benchmark ratios for the source fluxes
is $\Phi^0_e:\Phi^0_\mu:\Phi^0_\tau = 1:2:0$.
Lepton mixing changes this ratio to $\simeq 1:1:1$ at Earth~\cite{FR2}.
Depending on astrophysical processes at the sources or new physics  
involved in the production, propagation and detection,
the democratic composition at Earth may be significantly changed~\cite{FR1,FR2}. 

In this paper, we study the flavor composition of high-energy neutrinos 
by using the three years data of IceCube~\cite{IC3}.
Making the normalizations of power-law fluxes be flavor dependent,
we fit the data and report the best fit and the intervals for the normalizations.

This issue was first addressed in Ref.~\cite{Palomares-Ruiz}, where they found 
that the $1:1:1$ composition at Earth with $E_\nu^{-2}$ spectrum is disfavored 
at $92\%$ CL with the best fit composition $1:0:0$.
They analyzed the total number of the shower and the track events which are
integrated over the deposited energies.
A goal of this paper is to study the impact of the energy distributions
on the determination of the flavor ratios.
We model the astrophysical neutrino fluxes for each flavor $\Phi_\alpha\,(\alpha =
e,\mu,\tau)$ as $\Phi_\alpha \,=\, n_\alpha E_\nu^{-\gamma}$, where
$n_\alpha$ is the (flavor dependent) normalization, $E_\nu$ is the neutrino energy,
and $\gamma$ is the spectral index.
The model parameters to be determined are $n_\alpha$ and $\gamma$.
By seeking the global minimum of a $\chi^2$ function (see Section 3) with 
respect to these four parameters, we study the interplay between the flavor ratios 
and the spectral index.
Our emphasis is, however, not on the numbers themselves given by the analysis,
but on the qualitative differences between the flavored model and the usual democratic
model, which may be highlighted by taking account of the energy distribution.

This paper is organized as follows.
In Section~\ref{events}, the calculations for the number of events by the
astrophysical neutrinos are demonstrated.
In Section~\ref{flavor}, we discuss the method of the statistical analysis and 
show the results. 
Section~\ref{summary} is for conclusions.

%%%%%%%%%%%%%%%%%%%%%%%%%%%%%%%%%%%%%%%%%%%%%%%%%%%%%%%%%%%%%%%%%%%%%%
\section{Number of events}
\label{events}
%%%%%%%%%%%%%%%%%%%%%%%%%%%%%%%%%%%%%%%%%%%%%%%%%%%%%%%%%%%%%%%%%%%%%%
%%%%%%%%%%%%%%%%%%%%%%%%%%%%%%%%%%%%%%%%%%%%%%%%%%%%%%%%%%%%%%%%%%%%%%
\subsection{Astrophysical neutrino events}
%%%%%%%%%%%%%%%%%%%%%%%%%%%%%%%%%%%%%%%%%%%%%%%%%%%%%%%%%%%%%%%%%%%%%%
Following Ref.~\cite{IC2,IC3}, we focus on the neutrino events whose vertices are 
contained in the detector volume (so-called ``starting events'').
The neutrinos leave their signals via neutrino--nucleon ($\nu N$) scattering.
There are two main topologies of the neutrino events; the showers and the tracks.
The electron neutrinos $\nu_e$ trigger the shower events by 
the charged current (CC) and the neutral current (NC) interactions.
The muon neutrinos $\nu_\mu$ produce both tracks and showers by the CC and NC
interactions, respectively.
The tau neutrinos $\nu_\tau$ with the energies less than $\sim 1$ PeV produce
showers by CC and NC, whereas $\nu_\tau$ with energies greater than $1$ PeV may 
produce distinct events called double-bang and lolipop~\cite{tau}.
In this paper, we assume $\nu_\tau$ triggers only showers since we focus on the
neutrino events whose energies are less than a few PeV\footnote{The taus produced
from $\nu_\tau$-CC decay to muons in $17.4$\% branching ratio, and such events are 
classified as tracks.
The inclusion of this track events slightly changes the following results
on the flavor composition. 
However, the best fit values of the spectral index $\gamma$ are not changed.}.

Let us first focus on the down-going events where the attenuation by Earth
is irrelevant.
The number of the shower events by the CC interactions of $\nu_e N$ and $\nu_\tau N$  
are given by
\begin{eqnarray}
\nu^{\rm sh}_{\rm CC} \,=\,
2\pi T N_A  \!\int\!\! dE_\nu \,\, V_{\rm CC}^{e,\tau}\,\,
 \sigma_{\rm CC}\,\,\Phi_{e,\tau},
\label{rate1}
\end{eqnarray}
where $T=988$ days of exposure time,
$N_A = 6.022\times 10^{23}\,{\rm g^{-1}}$, $E_\nu$ is the neutrino energy, 
$V_{\rm CC}^{e}$ and $V_{\rm CC}^{\tau}$ are the effective masses of 
the detector~\cite{IC2},
$\sigma_{CC}$ is the $\nu N$ total cross section for the CC interactions~\cite{cross},
$\Phi_{e}$ and $\Phi_{\tau}$ stand for the $\nu_e$ and $\nu_\tau$ fluxes,
respectively.
The factor $2\pi$ accounts for the integration over Southern sky under the
assumption that the neutrino fluxes are isotropic.
In the CC channel of $\nu_e N$ and $\nu_\tau N$, almost all neutrino energy is
converted to the electromagnetic deposited energy ($E_{\rm em}$).
In what follows, we assume $E_\nu = E_{\rm em}$ for these CC processes.

The number of the shower events by the NC interactions of $\nu_\alpha N$ ($\alpha = 
e,\mu,\tau$) are given by
\begin{eqnarray}
\nu^{\rm sh}_{\rm NC} =
2\pi TN_A  \!\int_{E_0/\langle y \rangle}^{E_1/\langle y \rangle}\!\! dE_\nu \,\, 
V_{\rm NC} \,\,
 \sigma_{\rm NC}\,\,\Phi_{\alpha},
\label{rate2}
\end{eqnarray}
where $V_{\rm NC}$ is the effective mass for the NC processes~\cite{IC2},
$\sigma_{\rm NC}$ is the $\nu N$ total cross section for the NC 
interactions~\cite{cross},
and $\langle y \rangle$ is the mean inelasticity, which is the mean energy fraction 
carried by the kicked quark in the final state~\cite{cross}.
The formula Eq.~(\ref{rate2}) shows the number of events for the shower energy
between $E_0$ and $E_1$.
 
Finally, the number of the track events by the CC interactions of $\nu_\mu N$  
is given by
\begin{eqnarray}
\nu^{\rm tr} =
2\pi T N_A  \!\int_{E_0/\langle y \rangle}^{E_1/\langle y \rangle}\!\! 
dE_\nu \,\, V_{\rm CC}^{\mu}\,\,
 \sigma_{\rm CC}\,\,\Phi_{\mu},
\label{rate3}
\end{eqnarray}
where $V_{\rm CC}^{\mu}$ is the effective mass for the $\nu_\mu$ CC process~\cite{IC2}.
The out-going muons produced inside the instrumental volume usually escape from
the volume, such that the showers at the starting vertices dominantly contribute
to the deposited energies.
In this work, we assume the deposited energies are equal to the starting shower
energies, and use Eq.~(\ref{rate3}) for the track events whose deposited energies
between $E_0$ and $E_1$.

For the up-going events (the events induced by the neutrinos coming from Northern sky), 
the events are calculated by Eq.~(\ref{rate1}),~(\ref{rate2}) and (\ref{rate3})
with the replacement $(2\pi) \to (2\pi)S(E_\nu)$, where $S(E_\nu)$ is the
shadow factor~\cite{cross} varying from zero to unity, 
which accounts for the attenuation of the neutrinos by Earth.
The calculations for the antineutrino are done by replacing 
the cross-sections which are slightly different from the ordinary ones~\cite{cross}. 

%%%%%%%%%%%%%%%%%%%%%%%%%%%%%%%%%%%%%%%%%%%%%%%%%%%%%%%%%%%%%%%%%%%%%%
\begin{figure}[t]
\begin{center}
\scalebox{0.5}{\includegraphics{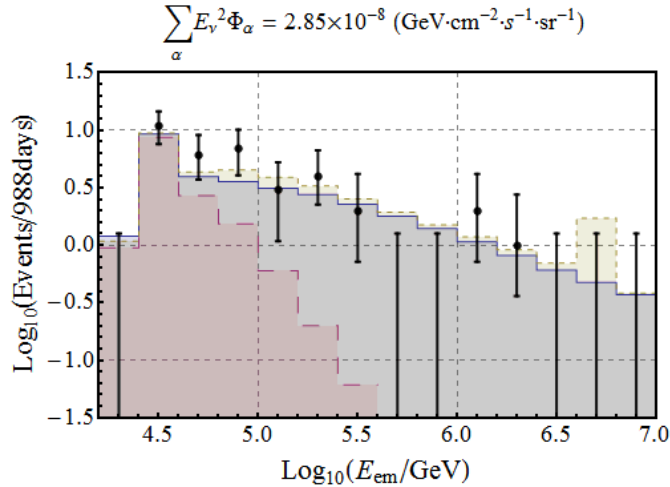}}
\end{center}
\caption{Deposited energy distribution of the astrophysical neutrino and
the background events.
The solid line shows the sum of the astrophysical neutrino events calculated by
Eqs.(\ref{rate1})--(\ref{flux1}) and the atmospheric background events (shown by the 
long-dashed line) simulated by the IceCube collaboration~\cite{IC3}.
The short-dashed line is the IceCube estimation of the total events
with $E^2 \Phi =  0.95\times 10^{-8}\,{\rm GeV\, cm^{-1}\, s^{-1}\, sr^{-1}}$
for each flavor. The black dots are the observed data.}
\label{fig1}
\end{figure}
%%%%%%%%%%%%%%%%%%%%%%%%%%%%%%%%%%%%%%%%%%%%%%%%%%%%%%%%%%%%%%%%%%%%%%

%%%%%%%%%%%%%%%%%%%%%%%%%%%%%%%%%%%%%%%%%%%%%%%%%%%%%%%%%%%%%%%%%%%%%%
\subsection{Astrophysical neutrino fluxes}
%%%%%%%%%%%%%%%%%%%%%%%%%%%%%%%%%%%%%%%%%%%%%%%%%%%%%%%%%%%%%%%%%%%%%%

In this work, we consider isotropic diffuse fluxes for the astrophysical neutrinos.
In order to make the model be sensitive to the neutrino flavors in a 
simple way, let the normalization of the astrophysical neutrino flux 
for each flavor be independent, while assuming the spectra follow a common 
power law;
\begin{eqnarray}
\Phi_\alpha \,=\, n_\alpha E_\nu^{-\gamma}, \quad\quad(\alpha = e,\mu,\tau).
\label{flux1}
\end{eqnarray}
Fig.~\ref{fig1}  shows typical examples of the deposited energy distributions of
the events.
The solid line shows the summation of the astrophysical neutrino events calculated by
Eqs.(\ref{rate1})--(\ref{flux1}) and the background events (shown by the  
long-dashed line) simulated by the IceCube collaboration~\cite{IC3}.
The IceCube estimation of the total events is also shown by
the short-dashed line for comparison. The black dots are the observed data.
In accordance with Ref.~\cite{IC3}, the flux for each flavor is set as 
$n_\alpha = 0.95\times 10^{-8}\,{\rm GeV\, cm^{-1}\, s^{-1}\, sr^{-1}}$
for $\alpha = e,\mu,\tau$ with $\gamma = 2.0$, and
the neutrino/antineutrino ($\nu/\bar{\nu}$) fraction is taken to be unity.
It is seen from Fig.~\ref{fig1} that the estimation by Eqs.(\ref{rate1})--(\ref{flux1}) 
agrees well with the IceCube analysis, up to the large discrepancy
at the bin for $E_{\rm em} = 10^{6.6}-10^{6.8}\,{\rm GeV}$.
The shower and the track fraction of the astrophysical neutrino events
are $82$\% and $18$\%, respectively. 
These numbers are  also agree with Ref.~\cite{IC3}.

The large discrepancy around $E_{\rm em} = 10^{6.7}\,{\rm GeV}$ is due to
the Glashow resonance~\cite{GR1}, which is the resonant production of the $W^-$ boson
in $\overline{\nu}_e e$ scattering at $E_\nu = 6.3\,{\rm PeV}$.
This effect is not included in Fig.~\ref{fig1}.
The significance of this resonance strongly depends on the $\nu/\bar{\nu}$
fraction~\cite{GR2}, which would be a nuisance to the current 
purpose.
Since no events larger than $\sim 2\,{\rm PeV}$ have been observed,
we first avoid the uncertainty from the $\nu/\bar{\nu}$ fraction 
by assuming that the power-law fluxes have a
cutoff at $E_\nu = 3.0\,{\rm PeV}$.
In this case, the $\nu/\bar{\nu}$ fraction does not make much difference 
on the result of the following flavor analysis.
The effect of the Glashow resonance on the fluxes without cutoff
is discussed later (see Table~\ref{tab2} and the related text).
In what follows, we set the $\nu/\bar{\nu}$ ratio to be unity as a typical
example. Such a ratio is realized if the neutrinos are produced on source 
by the proton-proton scattering.

%%%%%%%%%%%%%%%%%%%%%%%%%%%%%%%%%%%%%%%%%%%%%%%%%%%%%%%%%%%%%%%%%%%%%%
\section{Flavor compositions}
\label{flavor}
%%%%%%%%%%%%%%%%%%%%%%%%%%%%%%%%%%%%%%%%%%%%%%%%%%%%%%%%%%%%%%%%%%%%%%
We assume that the shower and track events are Poisson distributed around 
mean values $\mu^{\rm sh}$ and $\mu^{\rm tr}$.
The observed data is fitted by minimizing the logarithm of the likelihood ratio of
the current model to the saturated model~\cite{PDG}
\begin{eqnarray}
\chi^2 \,=\, \chi^2_{\rm shower} \,+\, \chi^2_{\rm track},
\label{chisq}
\end{eqnarray}
%where
\begin{eqnarray}
\chi^2_{\rm shower} \,=\, 2 \sum_{i=1}^{14}
\left(\, \mu^{\rm sh}_i - N^{\rm sh}_i + N^{\rm sh}_i\ln\frac{N^{\rm sh}_i}
{\mu^{\rm sh}_i}\,
 \right),
\end{eqnarray}
where $i$ labels the energy bins (see in Fig.~\ref{fig1}),
$N^{\rm sh}$ is the observed shower events~\cite{IC3}.
The mean of the shower events $\mu^{\rm sh}$ is given by
\begin{eqnarray}
\mu^{\rm sh} \,=\, \nu^{\rm sh} + b^{\rm sh},
\end{eqnarray}
where $\nu^{\rm sh}$ stands for $\nu^{\rm sh}_{\rm CC} + \nu^{\rm sh}_{\rm NC}$ 
summed over the up and down-going, the neutrino and antineutrino components.
$b^{\rm sh}$ is the background shower events.
The symbols with the subscript $i$ stand for the values for the $i$-th bin.
The function $\chi^2_{\rm track}$ is defined in the same manner as $\chi^2_{\rm shower}$.

For the background estimations of $b^{\rm sh}$ and $b^{\rm tr}$, we use
the numbers in Ref.~\cite{IC3}; the binned expected numbers for ``atmospheric 
neutrino $(\pi/K)$'' and ``muon flux'' shown in Fig.~2 of Ref.~\cite{IC3}.
In order to breakdown the atmospheric neutrino events into the showers and the tracks,
we assume that the atmospheric neutrino events are solely induced  by $\nu_\mu$ and its
CC and NC reactions are identified as the tracks and the showers, respectively.
This estimates that the track events account for $76\%$ of the 
atmospheric neutrino events in each energy bin\footnote{A more realistic number
given in Ref.~\cite{IC3} is $69$\%. If we use this number in the following
analysis, the best fit values of $\gamma$ and $n_\alpha$ are accordingly changed 
by a few percent.}. 

%%%%%%%%%%%%%%%%%%%%%%%%%%%%%%%%%%%%%%%%%%%%%%%%%%%%%%%%%%%%%%%%%%%%%%
\begin{figure}[t]
\begin{center}
\begin{tabular}{cc}
\includegraphics[scale=0.4]{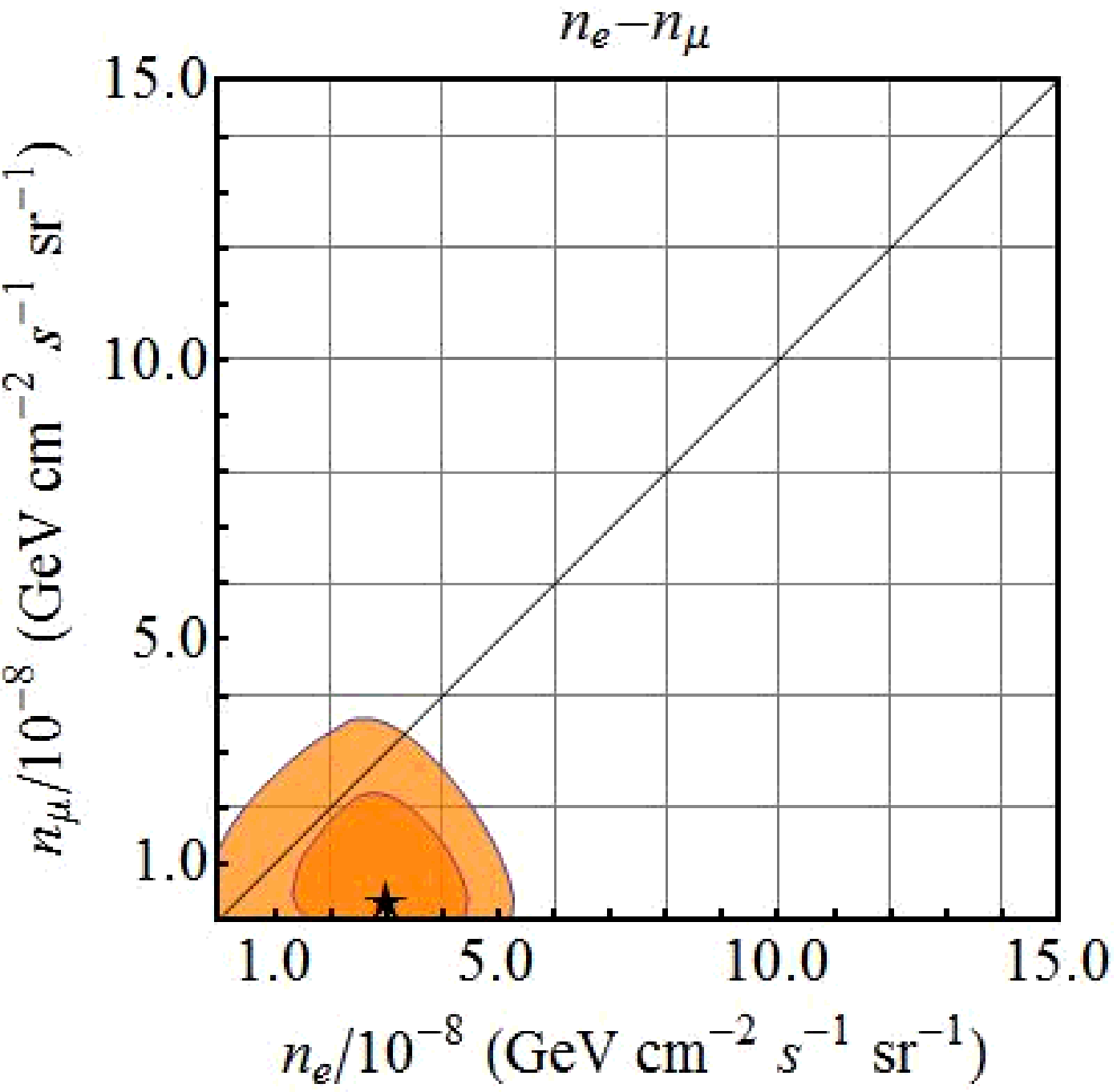}&  \\
\includegraphics[scale=0.4]{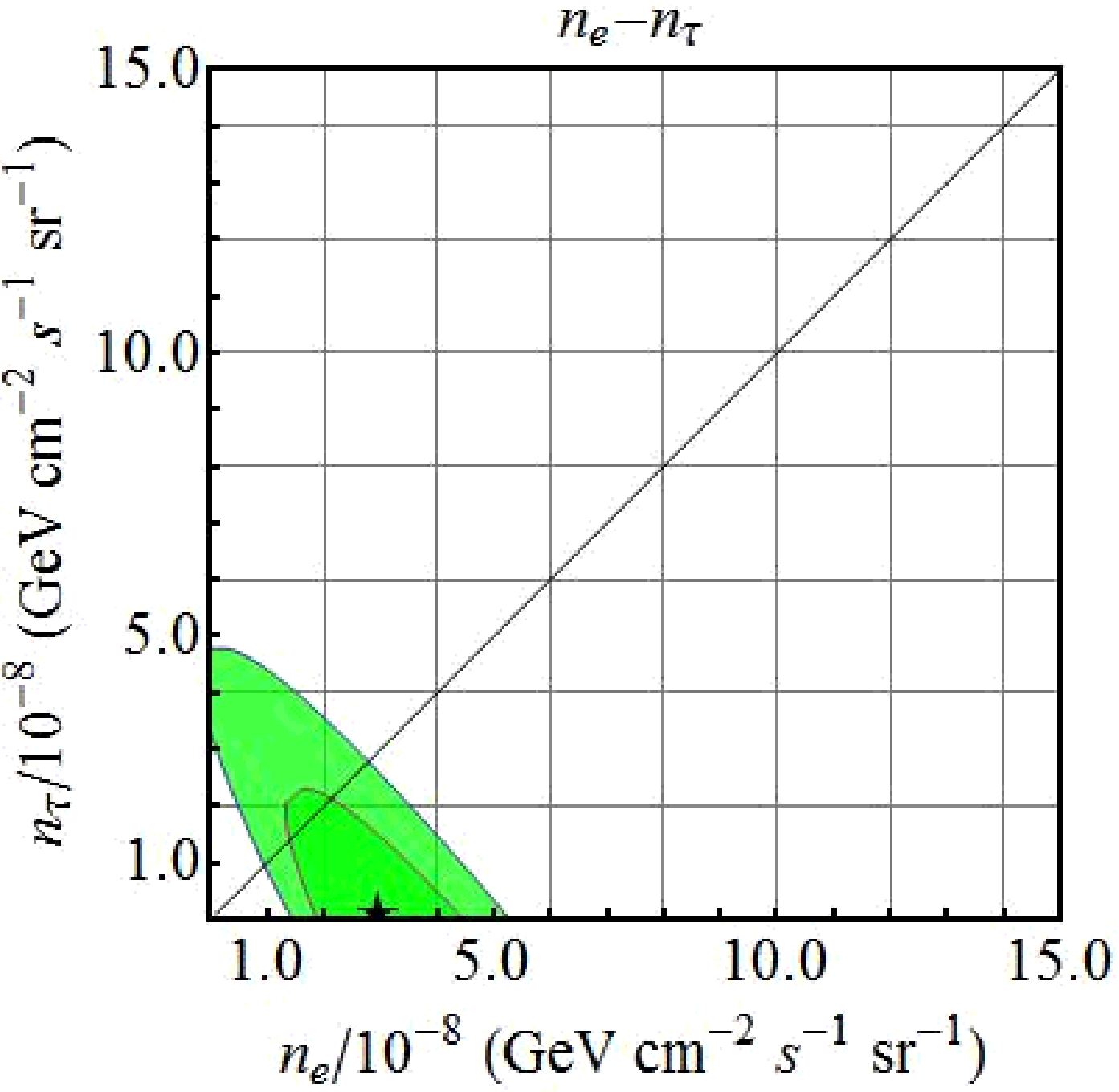}&\includegraphics[scale=0.4]{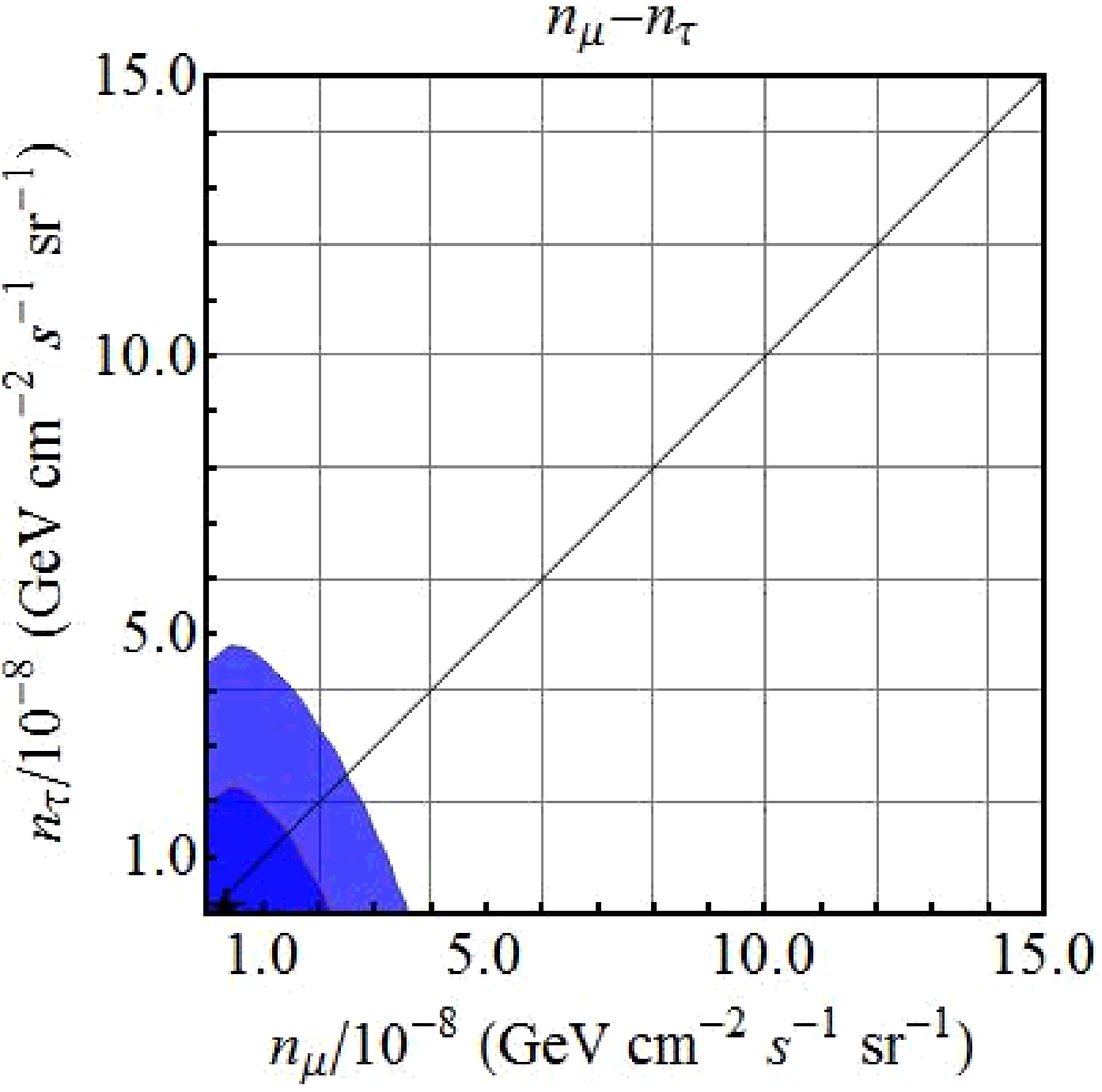}\\
\end{tabular}
\end{center}
\caption{Best fit and intervals of $n_e, n_\mu, n_\tau$ in the case of $E_\nu^{-2}$ 
spectrum ($\gamma = 2.0$).
The three panels show the regions in the three dimensional $(n_e,n_\mu,n_\tau)$ space
projected to the two dimensional planes.
The symbol $\star$ stands for the best fit, and the inner (outer) region
filled dark (light) is the 68\% (95\%) region.}
  \label{abc}
\end{figure}
%%%%%%%%%%%%%%%%%%%%%%%%%%%%%%%%%%%%%%%%%%%%%%%%%%%%%%%%%%%%%%%%%%%%%% 

For a fixed value of the spectral index $\gamma$, the best fit of 
$n_e$, $n_\mu$, $n_\tau$ is given by the minimum of Eq.~(\ref{chisq}).
In addition to the best fit, we report the regions which satisfy 
$\chi^2< \chi^2_{\rm min} + 3.53\,(7.82)$ as approximate 
68\% (95\%) confidence regions~\cite{PDG}.

Fig.~\ref{abc} shows the best fit and the intervals in the case of $E_\nu^{-2}$ 
spectrum ($\gamma = 2.0$).
The three panels show the projections of the regions in the three dimensional 
$(n_e,n_\mu,n_\tau)$ space.
The symbol $\star$ stands for the best fit, and the inner (outer) region
filled in dark (light) colors is the 68\% (95\%) region.
The best fit is $n_e =3.0\times 10^{-8}$, $n_\mu =3.9\times 10^{-9}$, 
$n_\tau = 0$ in the unit of ${\rm GeV \,cm^{-1}\, s^{-1}\, sr^{-1}}$
where $\chi^2_{\rm min} = 42.7/25\,{\rm dof}$.
Starting from the minimum, the $\chi^2$ function is well increasing along the $n_e$ 
axis, whereas it sharply stands up only toward the increasing direction
along the $n_\mu$ and $n_\tau$ axes.
Although the best fit of $n_\mu$ is not zero, the increasing of $\chi^2$ is  
moderate along the decreasing $n_\mu$ direction.

The standard $\Phi_e : \Phi_\mu : \Phi_\tau = 1:1:1$ hypothesis is represented
by the $n_e = n_\mu = n_\tau$ trajectory in the $(n_e, n_\mu, n_\tau)$ space.
It is seen from Fig.~\ref{abc} that $1:1:1$ is lying on outside 
of the $68\%$ region.
The minimum of $\chi^2$ along the  $n_e = n_\mu = n_\tau$ trajectory is
$\left. \chi^2_{\rm min}\right|_{n_e = n_\mu = n_\tau} = 46.9/27\,{\rm dof}$,
which means the $n_e = n_\mu = n_\tau$ trajectory is tangent to the $76\%$ surface
in the $(n_e, n_\mu, n_\tau)$ space.
If we change the background assumption by replacing the track fraction $76\%$ 
with $69\%$($50\%$), $\left. \chi^2_{\rm min}\right|_{n_e = n_\mu = n_\tau}$ 
goes down from $46.9$ to $44.5$($39.0$).

%%%%%%%%%%%%%%%%%%%%%%%%%%%%%%%%%%%%%%%%%%%%%%%%%%%%%%%%%%%%%%%%%%%%%%%
\begin{figure}[t]
\begin{center}
  \centerline{
  \includegraphics[scale=0.4]{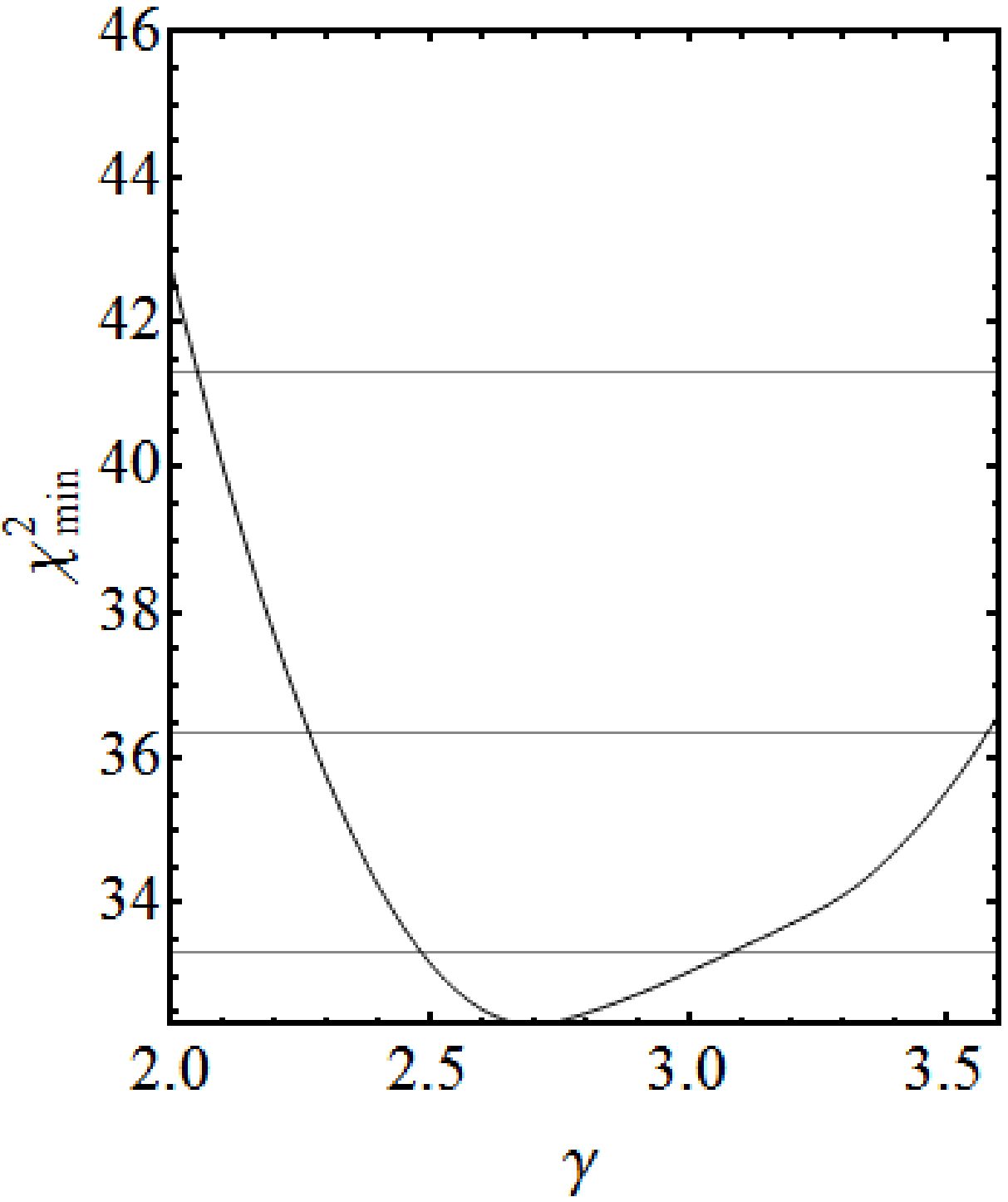}%
\hspace{5mm}
  \includegraphics[scale=0.4]{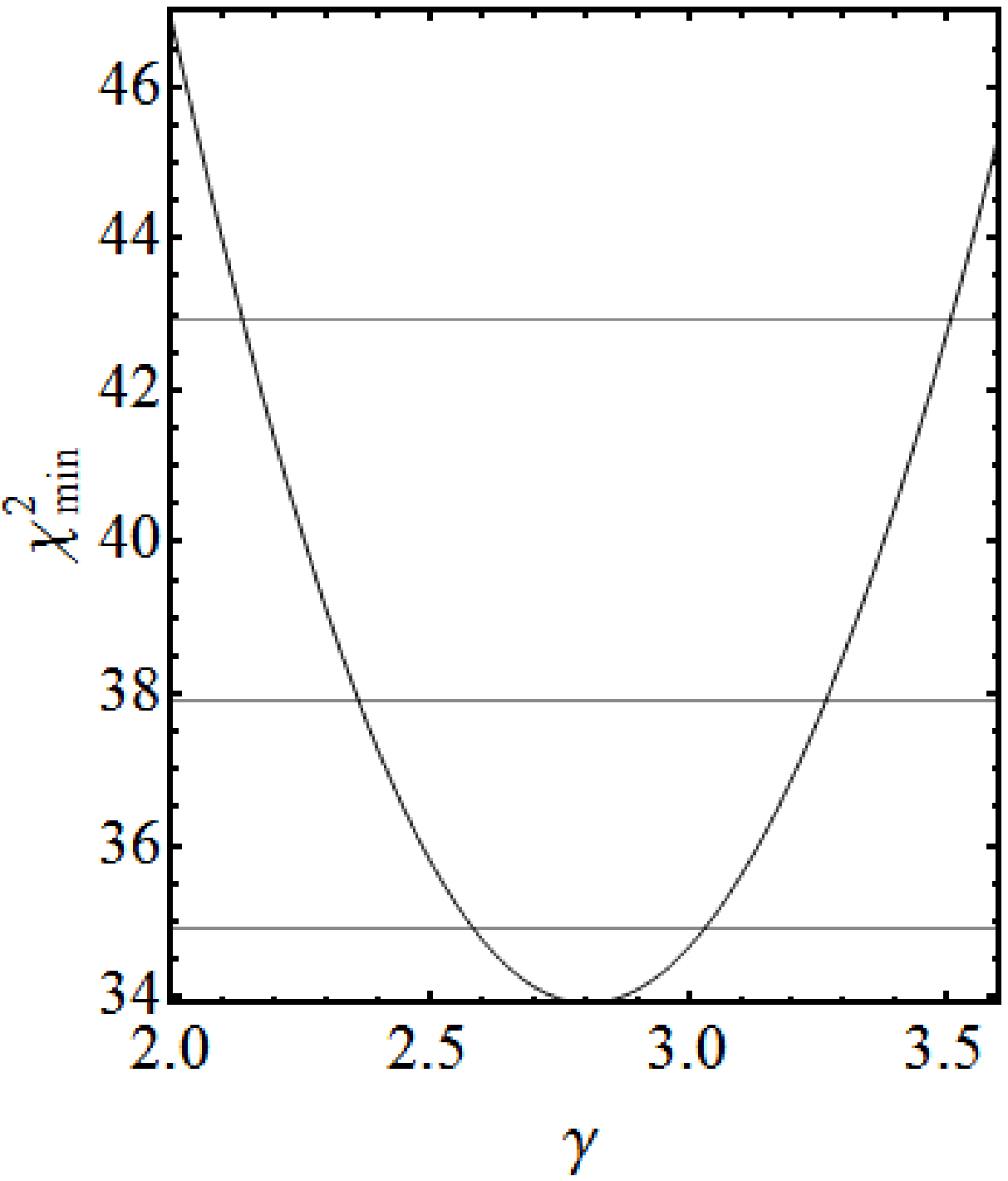}%
  }%
\end{center}
\caption{Minimum of $\chi^2$ (Eq.~(\ref{chisq})) for
each value of the spectral index $\gamma$.
The left panel shows the case where $n_e$, $n_\mu$ and $n_\tau$ are
independent, while the right panel shows
the case where the condition $n_e = n_\mu = n_\tau$ is imposed.
The horizontal lines show the minimum $+$ $1$, $4$, and $9$, 
as the references for the $1$,$2$ and $3\sigma$ ranges of $\gamma$.}
  \label{chigam}
\end{figure}
%%%%%%%%%%%%%%%%%%%%%%%%%%%%%%%%%%%%%%%%%%%%%%%%%%%%%%%%%%%%%%%%%%%%%%% 

The minimum of the flavored model with $E_\nu^{-2}$ spectrum is $\chi^2_{\rm min} = 
42.7/ 25\,{\rm dof}$,
which means that this fit must be also poor.
Better fits are obtained with the larger values of $\gamma$. 
Fig.~\ref{chigam} shows the minimum of $\chi^2$ (Eq.~(\ref{chisq})) for
each value of the spectral index $\gamma$.
The left panel shows the case where $n_e$, $n_\mu$ and $n_\tau$ are
independent, while the right panel shows
the case where the condition $n_e = n_\mu = n_\tau$ is imposed.
The left panel tells us that the global minimum is away from $\gamma = 2.0$.
The minimum is achieved at $\gamma = 2.7$, where 
$\chi^2_{\rm min} = 32.3/ 24\,{\rm dof}$, which is more acceptable than 
$\gamma = 2.0$.
In the right panel of Fig.~\ref{chigam}, $\chi^2_{\rm min} = 33.9/ 26\,{\rm dof}$ 
at $\gamma = 2.8$.
When we omit the events below $\sim 60\,{\rm TeV}$ and perform a fit without 
the lower three bins, the best fit index becomes 
$\gamma = 2.3$ for the flavored model and $\gamma = 2.4$ for the democratic model,
in agreement with Ref.~\cite{IC3}.

\begin{figure}[h]
\begin{center}
\begin{tabular}{cc}
\includegraphics[scale=0.4]{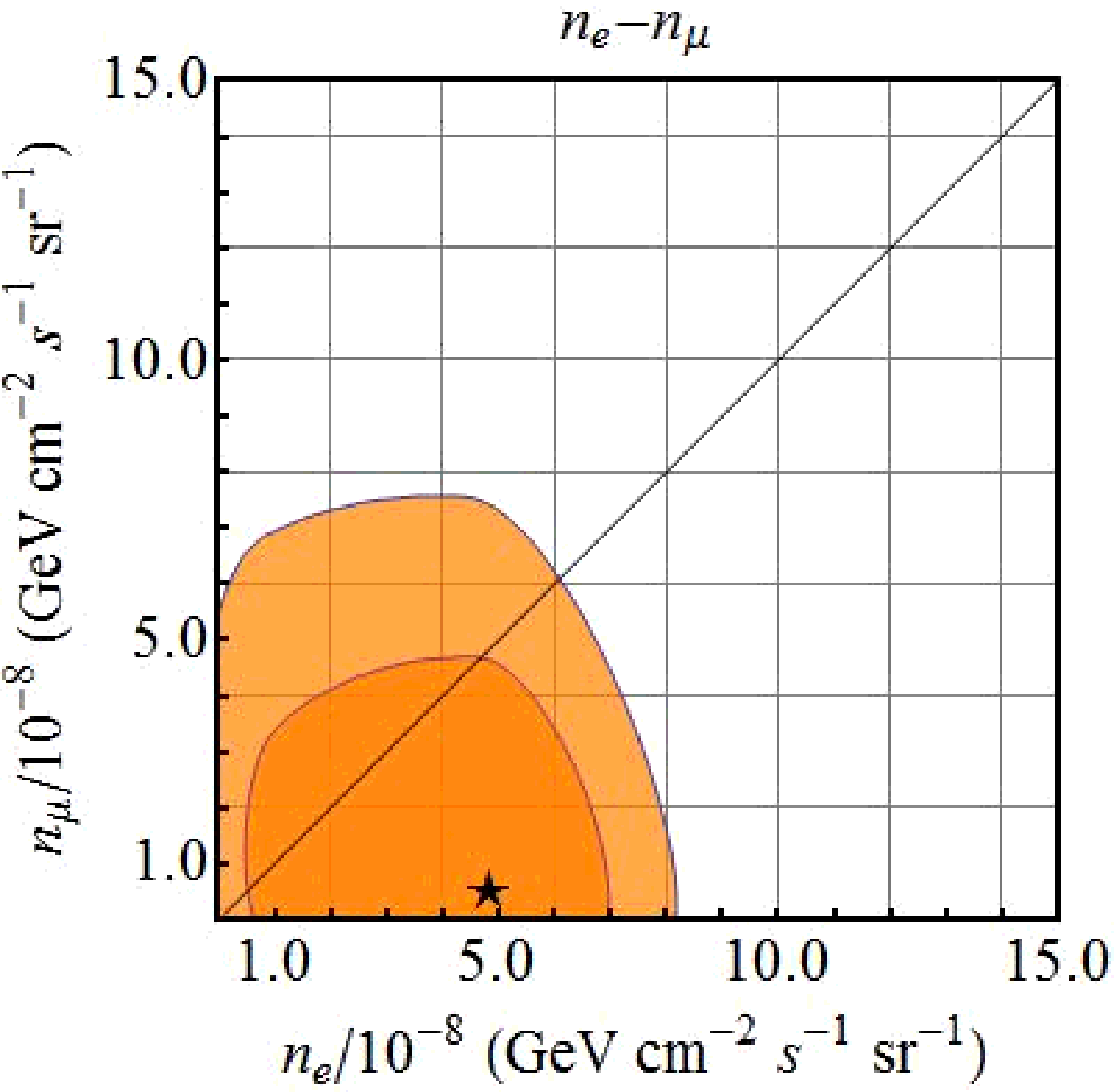}&  \\
\includegraphics[scale=0.4]{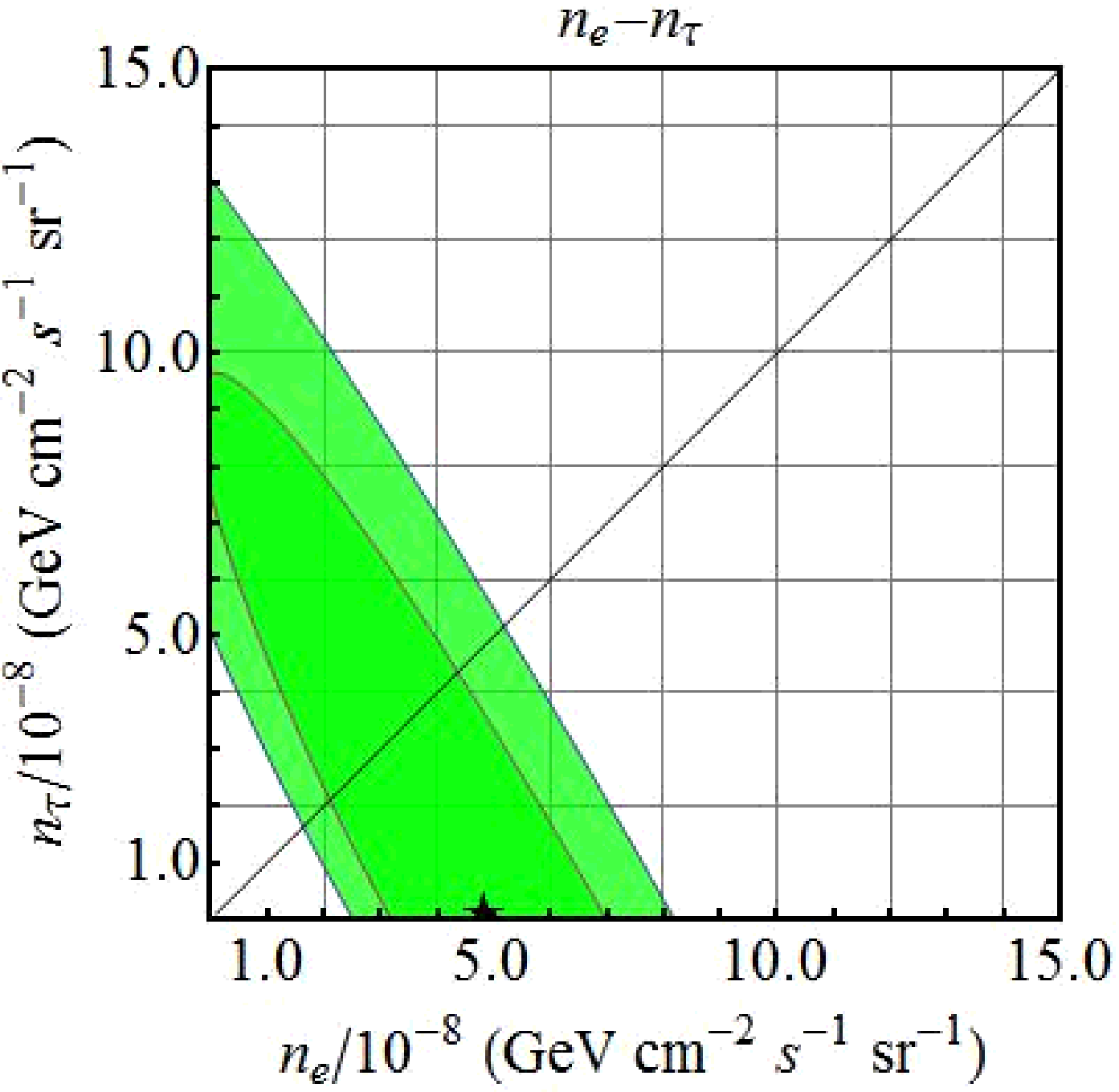}&\includegraphics[scale=0.4]{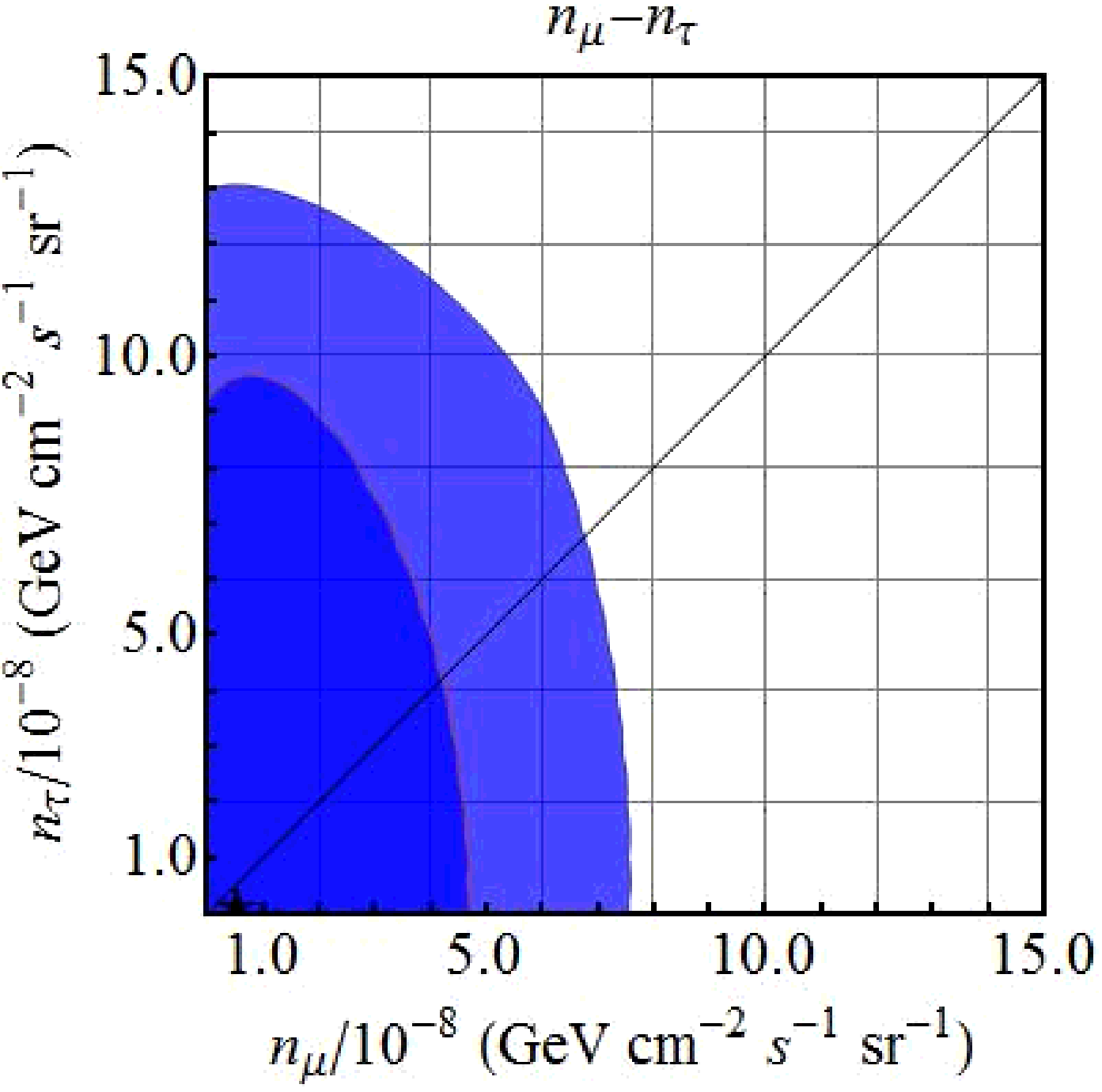}\\
\end{tabular}
\end{center}
\caption{
The same plots as in Fig.~\ref{abc}, but for $E_\nu^{-2.7}$ spectrum ($\gamma = 2.7$).
}
\label{abc2}
\end{figure}

Fig.~\ref{abc2} shows the regions for the normalization constants 
in the case of $\gamma = 2.7$.
In the plots, the normalization parameters are taken as
\begin{eqnarray}
E_\nu^2 \Phi_\alpha = n_\alpha \left( \frac{E_\nu}{10^5\,{\rm GeV}} \right)^{-0.7},
\quad\quad (\alpha = e,\mu,\tau).
\end{eqnarray}
The best fit is $n_e = 4.9\times 10^{-8}$, 
$n_\mu = 5.8\times 10^{-9}$, $n_\tau =0$ in the unit of 
${\rm GeV \,cm^{-1}\, s^{-1}\, sr^{-1}}$.
Compared with $\gamma = 2.0$ (Fig.~\ref{abc}), wider ranges are allowed
for $\gamma = 2.7$.
The $n_e = n_\mu = n_\tau$ trajectory is tangent to the $38\%$ surface, which
means the $1:1:1$ ratio is consistent with the current observation.
%%%%%%%%%%%%%%%%%%%%%%%%%%%%%%%%%%%%%%%%%%%%%%%%%%%%%%%%%%%%%%%%%%%%%%%%%
\begin{table}
\begin{center}
\begin{tabular}{c|c|c|c|c}\hline\hline
 & \multicolumn{2}{|c|}{$\gamma = 2.0$} & \multicolumn{2}{|c}{$\gamma = 2.7$} \\\hline
 & {\small 68\%} & {\small 95\%} & {\small 68\%} & {\small 95\%} \\ \hline
$T$ & $0$\,-\,$0.47$&$0$\,-\,$0.63$ &$0$\,-\,$0.53$&$0$\,-\,$0.70$ \\
$R$ & $0.62$\,-\,$\infty$ & $0$\,-\,$\infty$ &$0$\,-\,$\infty$ &$0$\,-\,$\infty$\\\hline
\end{tabular}
\end{center}
\caption{Crude intervals for the flux ratios 
$T = \Phi_\mu/(\Phi_e + \Phi_\mu + \Phi_\tau)$ and $R = \Phi_e/\Phi_\tau$.
In each column for $\gamma=2.0$ and $\gamma = 2.7$, the left (right)
item shows the interval corresponding to the 68\% (95\%) region
presented in Fig.~\ref{abc} and Fig.~\ref{abc2}.}
\label{tab1}
\end{table}
%%%%%%%%%%%%%%%%%%%%%%%%%%%%%%%%%%%%%%%%%%%%%%%%%%%%%%%%%%%%%%%%%%%%%%%%%

% It is shown in Ref. that about 30\% of the track events are misidentified as
% showers.

Note in passing that we are able to put the intervals on the flux ratios 
frequently quoted in literature.
The two ratios $T \equiv \Phi_\mu/(\Phi_e + \Phi_\mu + \Phi_\tau)$ and
$R \equiv \Phi_e/\Phi_\tau$ are often discussed~\cite{FR2}. 
As a crude estimate of the confidence intervals, we show 
in Table.~\ref{tab1} the ranges of the functions $T$ and $R$ under the
domain of the 68\%(95\%) regions of $(n_e, n_\mu, n_\tau)$ (the space defined by 
$\chi^2 \leq \chi^2_{\rm min} + 3.53\,(7.82)$).
Notice that this does not take into account the cancellation of the uncertainties,
so that the actual intervals may be narrower than shown here.

%%%%%%%%%%%%%%%%%%%%%%%%%%%%%%%%%%%%%%%%%%%%%%%%%%%%%%%%%%%%%%%%%%%%%% 
%%%%%%%%%%%%%%%%%%%%%%%%%%%%%%%%%%%%%%%%%%%%%%%%%%%%%%%%%%%%%%%%%%%%%% 

Finally, we comment on the effects of the Glashow resonance and 
the misidentification (mis-ID) of the track events.
It is pointed out that 30\% of the track events could be misidentified as 
showers~\cite{ICf}, 
and this effect has strong impacts on the determination of the flavor 
composition~\cite{Palomares-Ruiz2}.
In fact, we find that both of these effects have moderate impacts on the quality 
of the fit, but they significantly change the best fit of the flavor ratio 
and the exclusion level of $1:1:1$ in each model.

The results are summarized in Table~\ref{tab2}.
As is expected, the inclusion of mis-ID allows lager fractions of $\nu_\mu$ 
and reduces the exclusion level of
$1:1:1$.
In the flavored model with the four parameters $(\gamma, n_\alpha)$ being floated, 
the best fit ratio $1:0.1:0$ is changed to $1:0.2:0$, and the exclusion limit
38\% goes down to 12\%.
The best fit value of $\gamma$ is not changed by the mis-ID effects.

On the other hand, the effect of the Glashow resonance shifts the best fit of $\gamma$ 
to a slightly softer value, mitigating the conflict between the null observation
and the enhanced event rate at the resonance bin.
A striking change is the pusing-up of the $\nu_\tau$ component at the best fit.
There are two reasons for this increasing of $\nu_\tau$.
The first reason is the difference of the detection efficiencies of $\nu_e$ and 
$\nu_\tau$ at the lower energies. 
According to Ref.~\cite{IC2}, the effective volume of $\nu_e$ is 
as twice as large as $\nu_\tau$ around $40$-$100\,{\rm TeV}$.
As the spectrum gets soft, the events less than $\sim 100\,{\rm TeV}$ get
too large, so that $\nu_\tau$ is preferred for its lower detection rate than $\nu_e$.
In fact, in the search of the best fit ratio, $\nu_\tau$ becomes dominant over $\nu_e$
for $\gamma \gtrsim 2.8$ in any flavored model.
The second reason is that $\nu_\tau$ can account for the shower events
while keeping the resonance event suppressed. 
This effect may slightly push down the value of $\gamma$ at which $\nu_\tau$ overcomes 
$\nu_e$.
Since the inclusion of the Glashow resonance favors softer spectra, 
the best fit of the $\nu_\tau$ fraction accordingly increases due to the
first reason mentioned above.

%%%%%%%%%%%%%%%%%%%%%%%%%%%%%%%%%%%%%%%%%%%%%%%%%%%%%%%%%%%%%%%%%%%%%% 
\begin{table}[h]
\begin{center}
\begin{tabular}{cccc}\hline\hline
Model & Best fit & $\chi^2_{\rm min}$ & Exclusion level of 1:1:1\\\hline
%%%%%%%%%%%%%%%%%% Model 1%%%%%%%%%%%%%%%%%%%%%%%%%%%%%%%%%%%%%%%%%%%%%%%%%%%% 
$(\gamma, n_\alpha)$-free (4P)& 
\begin{tabular}{c}
$\gamma = 2.7$ \\
$1:\,0.1\,:0$  \\
\end{tabular} 
&  
\begin{tabular}{c}
$\chi^2_{\rm min} = 32.3/24 \,{\rm dof}$\\
\end{tabular}
& 38\% \\\hdashline
mis-ID & 
\begin{tabular}{c}
$\gamma = 2.7$ \\
$1:\,0.2\,:0$  \\
\end{tabular}
&
\begin{tabular}{c}
$\chi^2_{\rm min} = 32.2/24 \,{\rm dof}$\\
\end{tabular}
& 12\% \\\hdashline
GR & 
\begin{tabular}{c}
$\gamma = 2.9$ \\
$1:\,0.1\,:0.7$  \\
\end{tabular}
&
\begin{tabular}{c}
$\chi^2_{\rm min} = 32.9/24 \,{\rm dof}$\\
\end{tabular}
& 27\% \\\hdashline
mis-ID+GR & 
\begin{tabular}{c}
$\gamma = 2.8$ \\
$1:\,0.4\,:0.7$  \\
\end{tabular}
&
\begin{tabular}{c}
$\chi^2_{\rm min} = 32.8/24 \,{\rm dof}$\\
\end{tabular}
& 10\% \\\hline
%%%%%%%%%%%%%%%%%%% Model 2 %%%%%%%%%%%%%%%%%%%%%%%%%%%%%%%%%
$\gamma=2.0,\, n_\alpha$-free (3P)& 
\begin{tabular}{c}
\\
\\
\end{tabular}
\hspace{-5mm} 
\begin{tabular}{c}
$1:\,0.1\,:0$  \\
\end{tabular} 
&  
\begin{tabular}{c}
$\chi^2_{\rm min} = 42.7/25 \,{\rm dof}$\\
\end{tabular}
& 76\% \\\hdashline
mis-ID & 
\begin{tabular}{c}
\\
\\
\end{tabular}
\hspace{-5mm} 
\begin{tabular}{c}
%$\gamma = 2.7$ \\
$1:\,0.4\,:0$  \\
\end{tabular}
&
\begin{tabular}{c}
$\chi^2_{\rm min} = 42.2/25 \,{\rm dof}$\\
\end{tabular}
& 
47\% \\\hdashline
GR & 
\begin{tabular}{c}
\\
\\
\end{tabular} 
\hspace{-5mm}
\begin{tabular}{c}
%$\gamma = 2.7$ \\
$1:\,0.2\,:0.8$  \\
\end{tabular}
&
\begin{tabular}{c}
$\chi^2_{\rm min} = 54.1/25 \,{\rm dof}$\\
\end{tabular}
& 
32\% \\\hline
%%%%%%%%%%%%%%%%%%% Model 3 %%%%%%%%%%%%%%%%%%%%%%%%%%%%%%%%%
$(\gamma, n)$-free (2P)& 
\begin{tabular}{c}
\\
\\
\end{tabular}
\hspace{-5mm} 
\begin{tabular}{c}
$\gamma = 2.8$\\
%$1:\,0.1\,:0$  \\
\end{tabular} 
&  
\begin{tabular}{c}
$\chi^2_{\rm min} = 33.9/26 \,{\rm dof}$\\
\end{tabular}
& - \\\hdashline
mis-ID & 
\begin{tabular}{c}
\\
\\
\end{tabular}
\hspace{-5mm} 
\begin{tabular}{c}
$\gamma = 2.8$ \\
%$1:\,0.2\,:0$  \\
\end{tabular}
&
\begin{tabular}{c}
$\chi^2_{\rm min} = 32.8/26 \,{\rm dof}$\\
\end{tabular}
& 
- \\\hline
\end{tabular}
\caption{Summary of the best fit, $\chi^2_{\rm min}$, and the Feldman-Counsins 
exclusion level~\cite{FC} of $1:1:1$.
``$(\gamma, n_\alpha)$-free (4P)'' stands for the model where the four parameters
$\gamma, n_e, n_\mu, n_\tau$ are floated.
``$\gamma=2.0, n_\alpha$-free (3P)'' is the model where three parameters
$n_e,n_\mu,n_\tau$ are free with the fixed index $\gamma = 2.0$.
The third model ``$(\gamma,n)$-free (2P)'' is the case where 
$\gamma$ and the normalization $n=n_e=n_\mu=n_\tau$ are varied.
The sub-items ``mis-ID'', ``GR'', and ``mis-ID+GR'' show
the options that include the effect of the 30\% misidentification of
the tracks as showers, the Glashow Resonance without the energy cutoff,
and the combination of both, respectively.
 In the column of ``Best fit'', the ratios shows  $1:n_\mu/n_e: n_\tau/n_e$
at the best fit values of the normalizations.
}
\label{tab2}
\end{center}
\end{table}

%%%%%%%%%%%%%%%%%%%%%%%%%%%%%%%%%%%%%%%%%%%%%%%%%%%%%%%%%%%%%%%%%%%%%% 

%%%%%%%%%%%%%%%%%%%%%%%%%%%%%%%%%%%%%%%%%%%%%%%%%%%%%%%%%%%%%%%%%%%%%%
\section{Conclusions}
\label{summary}
%%%%%%%%%%%%%%%%%%%%%%%%%%%%%%%%%%%%%%%%%%%%%%%%%%%%%%%%%%%%%%%%%%%%%%
The current data of the IceCube's starting events seemingly shows a paucity of the
muon events. Above $30\,{\rm TeV}$, just eight tracks have been observed against
the background of $8.4\pm 4.2$ cosmic ray muon events and $6.6^{+5.9}_{-1.6}$ 
atmospheric neutrino events.
If this tendency would hold, it suggests that the standard $1:1:1$ scenarios
should be revised, and may even indicate the existence of some new physics.

In this work, we have studied the flavor composition of the astrophysical neutrinos 
observed in IceCube, especially focusing on the impact of the spectral index $\gamma$. 
Our point is not to give a precise estimation for the best fit and the intervals of
the relevant parameters, but to illustrate important qualitative features 
in the flavor and the spectrum analysis of the astrophysical neutrinos.
For this purpose, we consider the model with the three-flavor normalizations 
$n_e$,$n_\mu$,$n_\tau$ and a common index $\gamma$ kept independent (the flavored model), 
and compare it to the usual model with a common normalization and 
an index (the democratic model).

It is found that the global minimum of the flavored model is at $\gamma = 2.7$ 
with $\chi^2_{\rm min} = 32.3/24\,{\rm dof}$.
As for the democratic model, the best fit is at $\gamma = 2.8$ 
with $\chi^2_{\rm min} = 33.9/26\,{\rm dof}$.
The democratic model and the flavored model do not have much difference
in the quality of the (energy-spectrum) fit.
The standard $1:1:1$ composition is consistent with the current data.

However, the flavor compositon may affect the interval determination of $\gamma$.
The left panel of Fig.~\ref{chigam} shows that the $\chi^2$ does not
quickly stand up as $\gamma$ increases, indicating that the determination
of $\gamma$ might be more challenging for the flavored model than for the 
democratic case. The current background model does not leave much room
for the track contributions from the astrophysical neutrinos at lower energies.
Thus the $1:1:1$ case gets trouble at the lower energy bins as $\gamma$ becomes large, 
whereas the flavored model can avoid the conflict by taking the configuration
where the muon component is suppressed.
The inference of the spectral index may become a nontrivial task once the flavor
degress of freedom are switched on.

%%%%%%%%%%%%%%%%%%%%%%%%%%%%%%%%%%%%%%%%%%%%%%%%%%%%%%%%%%%%%%%%%%%%%%
\subsection*{Acknowledgments}
I thank Werner Rodejohann for his contribution in the early stage
of this work. 
I also thank Thomas Schwetz and Hiroaki Sugiyama for useful discussions on statistics.
\bigskip
%%%%%%%%%%%%%%%%%%%%%%%%%%%%%%%%%%%%%%%%%%%%%%%%%%%%%%%%%%%%%%%%%%%%%%

%%%%%%%%%%%%%%%%%%%%%%%%%%%%%%%%%%%%%%%%%%%%%%%%%%%%%%%%%%%% 

\end{document}